# ATRIUM - Architecting Under Uncertainty

## for ISO 26262 compliance

Naveen Mohan *, Per Roos †, Johan Svahn †, Martin Törngren *, Sagar Behere*
Corresponding author email: naveenm@kth.se
*KTH Royal Institute of Technology, Stockholm, Sweden
†Scania CV AB, Södertälje, Sweden

*Abstract*—The ISO 26262 is currently the dominant standard for assuring functional safety of electrical and electronic systems in the automotive industry. The Functional Safety Concept (FSC) sub-phase in the standard requires the Preliminary Architectural Assumptions (PAA) for allocation of functional safety requirements (FSRs).

This paper justifies the need for, and defines a process *ATRIUM*, for consistent design of the PAA. *ATRIUM* is subsequently applied in an industrial case study for a function enabling highly automated driving at one of the largest heavy vehicle manufacturers in Europe, Scania CV AB. The findings from this study, which contributed to *ATRIUM*'s institutionalization at Scania, are presented.

The benefits of the proposed process include (i) a fast and flexible way to refine the PAA, and a framework to (ii) incorporate information from legacy systems into safety design and (iii) rigorously track and document the assumptions and rationale behind architectural decisions under uncertain information.

The contributions of this paper are the (i) analysis of the problem (ii) the process *ATRIUM* and (iii) findings and the discussion from the case study at Scania.

*Keywords—ISO 26262, functional safety, automation, HCV, HGV, architectures, highly automated driving, ATRIUM, decision making, architecting, uncertainty management*

## I. Introduction

The modern vehicle can have upwards of a 100 Electronic Control Units (ECUs) and 300 functions distributed across these ECUs. Designing safety critical systems that perform in such a complex environment can be daunting. The ISO 26262 [1] was designed with this in mind to address functional safety, and provides a reference lifecycle to identify and mitigate risk in vehicular functions. The standard prescribes a top down approach for development of safety critical systems that, while ideal for development of a new system, poses significant challenges for the traditional automotive industry practice of reusing existing elements for cost efficiency [2].

A top-down analysis and breakdown of a function without consideration of the platform elements (that need to be reused anyway due to cost considerations) will lead to an inordinate amount of rework and expense to meet the safety goals in the later stages of the project. Thus, most architects and safety engineers do keep platform considerations in mind while designing functions even if it is without explicitly articulating them. This raises the issue that while architects are knowledgeable about the functions of the vehicle as a whole, it is not humanly possible to have detailed knowledge of the entire automotive platform. The general nature of architecting also implies that uncertainty in information is common and architectural decisions may be made based on potentially incomplete information. Architecting the "right" solution depends heavily on the expertise of those involved in the decisions and their implicit knowledge. As the industry heads towards automated driving and vehicular systems take over more control from the driver, the knowledge needed for design becomes more diverse. It is therefore a necessity from the industrial perspective to manage uncertain and implicit information in as explicit a way as possible to track the effect of this information on architectural design elements, while documenting rationale behind decisions. A systematic process to handle uncertainty will enable not only better traceability, but also consistency and repeatability in architectural decisions. Thus, future changes due to new information will become easier to manage.

This paper aims to answer the following questions:

**How can the decision making of architects under uncertain information be best managed to achieve the traceability needed for design of safety critical systems?**

and

**How can the knowledge from legacy systems be used to improve the design of the Preliminary Architectural Assumptions (PAA), as defined by the ISO 26262?**

We present the relevant background in section II and analyze the problem in Section III. In Section IV, we describe the ontology for, and propose the "ArchiTectural Refinement using Uncertainty Management" (*ATRIUM*) process to handle decision making under uncertainty and to imbibe knowledge of platform elements into a systematically developed and refined PAA. In Section V, we describe the experiences in the application of *ATRIUM* in an automotive case study and discuss the results. Section VI evaluates the method and presents its limitations before Section VII's related works and the conclusions in Section VIII.

## II. BACKGROUND

### A. A Short Note on Terms

Architecture is defined by ISO 42010 as" *fundamental concepts or properties of a system in its environment embodied in its elements, relationships, and in the principles of its design and evolution"* [3]. The ISO 26262 however takes a more tangible definition of the architecture as the *"representation of the structure of the item (1.69) or functions or systems (1.129) or elements (1.32) that allows identification of building blocks, their boundaries and interfaces, and includes the allocation (1.1) of functions to hardware and software elements 1.4"* [1]. The PAA though not formally defined in the ISO 26262 is understood to be the collection of knowledge about architecture in question, which is defined by the ISO 26262 to be comprised of *elements* which can be hardware, software or functional in nature. The PAA can contain information of a broad variety regarding any known information about the architecture used in the concept phase of the project i.e. the Preliminary Architecture (PA).

We use the terms method, process and tool as defined by Estefan in [4] throughout this paper. A *"Process (P) is a logical sequence of tasks performed to achieve a particular objective. A process defines "WHAT" is to be done, without specifying "HOW" each task is performed."* In addition, a method *"consists of techniques for performing a task, in other words, it defines the "HOW" of each task. A process comprises of tasks which performed by methods which are processes in their own right".* [4]

The terms L0, 1..5 arise from the SAE standard J3016 [5] which defines six levels of automation in driving, from Level 0(no automation) to L5 (full automation under all driving modes).

Through this paper, the term *architects* refer to architects at the vehicular level, unless specified otherwise. The other terms used for organizational roles are *safety engineers* who are tasked with assuring functional safety of the product and *experts* who are the experts in specific areas and technologies.

### B. ISO 26262

The ISO 26262, is an automotive specific adaptation for the generic standard for functional safety in electronics, the IEC 61508 [6], and gives a framework for the entire safety lifecycle, from concept phase up until the decommissioning of the system. The effectiveness of the standard relies on generating enough evidence and making arguments that make up the safety case for the function in question. To do so, the standard mandates a strict mapping of safety requirements to architectural elements. Functional Safety Requirements (FSRs) are created and allocated to architectural elements as early as the concept phase as part of the functional safety concept (FSC) sub-phase. The standard then provides a framework and strict rules to refine, breakdown and trace these requirements down to HW and SW implementation. The strength of the standard relies on the ability to trace the risk of a hazard at a vehicle level down to the implementation and guide the measures to mitigate the risk associated with the hazard at each level. The standard thus demands a high level of traceability through all the refinement of the FSRs and hence implicitly the PA.

This paper primarily deals with the FSC sub-phase in the concept phase. The objective of the FSC *"is to derive the functional safety requirements, from the safety goals, and to allocate them to the preliminary architectural elements of the item, or to external measures"* [1] . The standard places significance on the PAA by requiring it for deriving functional safety requirements by stating, *"The functional safety requirements shall be derived from the safety goals and safe states, taking into account the preliminary architectural assumptions".* The dependence of FSC on the PAA implies that an incorrect or incomplete PAA will lead to expensive iterations of the safety concept as corrections are made. Thus, the PAA should be as refined as possible to mitigate these architectural risks early on.

## III. ANALYSIS OF THE PROBLEM

Architecting in the cost-sensitive automotive domain is a multidimensional process. Other than technical knowledge, an architect also simultaneously considers several different design constraints including legal requirements, cost, variability, limitations from different target markets, organizational competence in technologies, marketing, strategic supplier relationships etc. Legacy systems and the need to reuse platform elements to optimize for cost, also influence architectural decisions. Automotive OEMs also have many tiers of technology suppliers and architects have to account for limitations based on integration of systems that were not designed for purpose. Due to this complex nature of automotive architecting, there is generally no unambiguously good architecture. There are always trade-offs and new properties are always discovered in the later stages of development that may influence previous decisions.

Though many tools to support architecting have been proposed, industrial adoption has been slow. Manual analysis methods are frequently used in the industry [7] [2] [8] and decisions are generally made by consensus amongst a group of architects and stakeholders. Depending on the makeup of a particular group, in terms of experience in the domain and their collective expertise in the various technologies used in the function, different results could be obtained. A way used in the industry to limit this variance is to choose the grouping appropriately and to involve specialists in the decision making process.

A degree of flexibility is unavoidable while working with architecture, in stark contrast with the needed rigid-ness and rigour of safety engineering. Incorrect architectural decisions become architectural risks that not only cause expensive changes, but could also compromise the safety argumentations.

As an example of the decisions that architects have to make under uncertain information, consider a function of L3 level of automation designed to exercise lateral and longitudinal control in a highway like scenario, once the driver has handed the function the control of the vehicle. The driver has been guaranteed 10 seconds of leadtime before the system relinquishes control, and is allowed to perform other tasks i.e. does not need to constantly supervise the dynamic driving task. One of the primary sensors used for the localization in this hypothetical scenario is the Forward Looking Camera (FLC), which is used to detect lane markings and is used in conjunction with a GPS sensor and controller to keep the truck within the lane. A forward facing radar and 4 side facing radars are used to track objects in the immediate vicinity. Staying within the lane has been identified as a key necessity for ensuring safety while the function is in control of the vehicle.

The task of the architect is then to assure that the architecture is capable enough to allow the function to operate in a safe albeit degraded capacity during the 10s even under failures. Consider a failure mode where the FLC becomes unavailable e.g. due to a blown fuse during operation, leading to a partial loss of localization. There could be many architectural design alternatives to ensure degraded operation in those 10s.

The architect could
a) change the existing GPS sensor to a high accuracy GPS sensor and use information about the road such as width, number of lanes etc. to stay within the lane.
b) have the function rely on features of the road, such as sound barriers, being detected by the side facing radars to stay within the lane.
c) always ensure that the truck is following another vehicle on the road such that it is always possible approximate its position on road w.r.t. the lead vehicle.
d) add HD maps and localization based on e.g. LIDARs for redundancy
  etc.

Each of these design alternatives has their merits and weaknesses. In terms of cost, a) and d) can be very expensive. c) limits the operational envelope of the function negatively, while b) though potentially cheaper, suffers from a lack of accuracy and limited applicability. The choice of the right solution depends on the factors discussed in the beginning of this section. However, irrespective of the specific selection, each of these alternatives carries risks due to them being untested and the absence of detailed behavioural information at the early stages of the project. The risks may be technical as discussed above or non-technical e.g. confidence in the ability of a supplier to deliver a production ready ECU before start of series production, cost of the product, lack of resources within the OEM to maintain a new technology etc..

The wide range and complexity of automated functions will require the architects to make several such decisions when the toy example is scaled up to that of a complete vehicle and more of the nuances of the platform needs to be considered. *ATRIUM* is designed to allow the architects to decide and trace the impact of their decisions for scenarios like this.

The following sources of architectural risks and uncertainty in information used to design the PAA were discovered during the analysis of literature (Section 7), and from interviews with system architects.

- Dynamic constraints**:** Operational scenarios and functional requirements are subject to change during the product lifecycle as trade-offs are determined.

- Functional allocation: Functional allocation to hardware is subject to business needs e.g. modularity and maintaining product lines for different markets.

- Technology immaturity: The typical lifetime of a new project in an automotive OEM is about three years. Failure modes that are implementation dependent and information about these are not available during concept design and may lead to changes later.

- Variability mapping: An element in a platform may have multiple variants with only a subset fulfilling the needed requirements. Lower limits on element specifications, for variant qualification, are thus needed, but these vary from case to case.

- Inconsistency in metrics: The weightage or relative importance of architectural metrics could change depending on situation.

- Pattern based design: The use of architectural patterns might determine selection of a difficult solution instead of a simpler one for the sake of consistency.

- Strategic relationships: Business needs might necessitate cooperating with other OEMs in specific areas for joint development, causing a non-optimal solution to be chosen.

- Dependencies on suppliers: Automotive suppliers cater to many different OEMs and purchased supplier systems often have unforeseen dependencies if they are not designed specifically for the organization in question.

- Roadmap related: Planned introduction of other related functions influence the selection of a suboptimal solution to have more optimal solutions in the future. Functions are not possible to be studied and judged in isolation.

- Rationale management: Rationale management is an essential in ensuring that tacit knowledge is made explicit, but is not usually practiced.

- Distributed expertise: Architects do not have detailed information about all of the existing platform elements. Particularly with the advent of driving functions with a high degree of automation, specialist information from a greater number of systems is needed and optimal solutions for the whole platform are harder to find.

- Failure modes management: A systematic way for finding out all failure modes of elements is not currently available.

- Insufficient change management: There is a lack of tool support for tracing and management of changes effectively.

- Lack of details in requirements: FSRs usually lack details needed to judge which elements can be used to completely fulfil them

- Change management of elements from ISO 26262: Architectural elements may change during design. Change management of these elements is not directly addressed in the ISO 26262.

- Incomplete definitions: PAA is not formally defined in ISO 26262 and hence the content and the level of abstraction needed are not understood.

While this is not a comprehensive list, it does become clear that architectural risks cannot be entirely avoided. However, the impact of these could be mitigated by managing the uncertainty explicitly. Overall uncertainty can also be reduced by incorporating knowledge from legacy systems that are required to be reused due to cost issues. We thus argue for incorporating as much information as available into early design phases and using stringent traceability to isolate and identify the changes uncertain information bring to the design. This allows for not only managing uncertainty caused by architectural decisions, but also for a consistent architectural design process and faster impact analysis on changed information.

## IV. THE ATRIUM PROCESS

This section describes the design and the ontology of the "ArchiTectural RefInement using Uncertainty Management" (*ATRIUM*) process.

### A. Requirements

Analysis of the problem in section 3, examination of the literature, workshops with intended users and stakeholders identified at Scania and multiple iterations on toy examples generated a set of requirements that *ATRIUM* needed to fulfil in order to answer the questions posed in section 1. These requirements are found in Table 1. The primary users were identified as the architects while the secondary users were defined to be the safety engineers and the experts (both from within and out of the organization).

TABLE I. REQUIREMENTS USED IN THE DESIGN OF *ATRIUM*

| # | Requirement text |
|---|---|
| R1 | The process shall document rationale behind architectural decisions. |
| R2 | The process shall follow the principles of the ISO 26262's change management strategy as far as feasible in the tracking of changes |
| R3 | The process shall facilitate the independence in the working between the primary and the secondary users for distributed development |
| R4 | The process shall allow information to be traceable to design decisions and specific elements in the architecture |
| R5 | The process shall facilitate refinement of PAA according to the decided failure modes and behaviours |
| R6 | The process shall facilitate limiting rework needed to only elements affected by specific information as far as possible. |
| R7 | The process shall allow multiple design solutions to be analysed simultaneously. |
| R8 | The process shall be designed in iterations to allow for baselines and intermediate solutions during design |
| R9 | The process shall be able to handle new elements being added into the PA within the same iteration |
| R10 | The process shall limit rework caused by addition of new failure modes in a new iteration. |
| R11 | Each iteration shall take the output of the previous iteration into consideration and lead to a more refined architecture and/or reduce uncertainty about the architectural risks involved |
| R12 | The process shall not be dependent on any specific organization. |
| R13 | The process shall be independent to the nature of the failure modes being analysed. |
| R14 | The process shall be independent to the nature of the elements being analysed. |
| R15 | The process shall not impose specific methods to decide between architectural alternatives. |
| R16 | The process shall make it possible to address multiple variants of an element. |
| R17 | The process shall document the architectural risks for the output and assign tasks to mitigate these risks. |
| R18 | The process shall use an inductive analysis technique such as FMEA to investigate the functional effects for all combinations of failure modes and elements |

### B. Process and Ontology

This sub-section gives a short explanation about *ATRIUM*, designed to fulfil the requirements in section 4.1, and the ontology used. The overall flow of *ATRIUM* is shown using an UML activity diagram in Fig. 1. Due to space limitations, many of the activities of the process are abstracted away with representative names. The most im-portant activities, marked in dark blue, are however expanded in Fig 2 and Fig 3. The groupings in Fig 1, marked with step-group1...4 and depicted with boxes having a dashed-line background are merely logical groupings to assist in discussion in Section 5. The terms used in the context of *ATRIUM* are based on natural language to facilitate understanding and *italicized* in this paper to distinguish from any other usage.

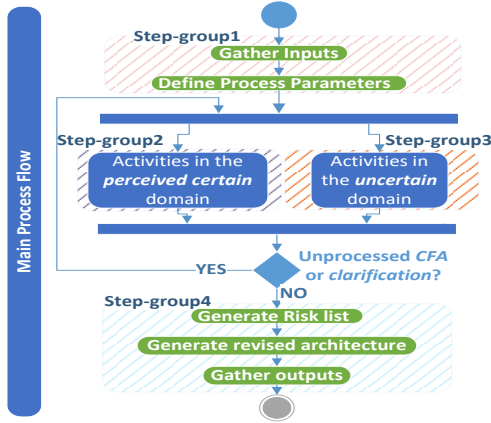

Fig. 1. *ATRIUM* overview

An *iteration* represents one complete execution of the process as shown in Fig. 1. The core of *ATRIUM* involves separation of available information into two domains namely the *perceived certain* and the *uncertain* domain. The information in *perceived certain* domain is static and traceability and consistency can be maintained. The *uncertain* domain contains information that could be subject to change. *Assumptions* serve as the meeting point between these two domains and transform *uncertain* information into *perceived certain* information. An *assumption* may correspond to a functional goal, constraint, operational condition or any information needed for architectural decisions. The nature of *assumption* is deliberately kept generic to accommodate for the wide variety of inputs that are used by architects. An *assumption* has a single qualifier of *validity*, which can take values of *valid* or *invalid*. *Assumptions* can be added at any point during the process flow. In *ATRIUM* Architects primarily work in the *perceived certain* domain while the other stakeholders i.e. the experts and the safety engineers work in the *uncertain* domain. The experts provide the service of clarifying and correcting information in the *uncertain* domain for the architects.

An *element* has the same meaning as in the ISO 26262 as discussed in section 2.2. An *element* has a qualifier called *state*. *State* can take values of legacy or new development. *Legacy* is assigned as value for the *element* if it has been available before the first iteration of the process is initiated. If the *element* is created during the execution of *ATRIUM*, *state* is assigned the value of *new*. The platform thus is made up of *legacy elements* and *new elements* are added to it with *ATRIUM*.

A *Component Failure Alternative* or *CFA* is a unique combination of a failure mode and an *element*. Each *CFA* has a single qualifier *state*, which can have either of the two values, *processed* or *unprocessed*. A *processed* value indicates that the *CFA* in question has been analysed to the best of the current knowledge available within the process iteration. An *unprocessed CFA* indicates that either the *CFA* was never analysed or that new information at least partially invalidated the *CFA* analysis.

The inputs referred to by the "gather inputs" abstraction in Fig 1. refer to the output of the previous iteration (if available), the technology roadmap of the organization and information about the *elements*.

A *Design Goal* or *DG* is the intended behaviour that the vehicle should achieve in case of a failure. Each *CFA* is linked to exactly one *DG*. A *DG* is comprised of one or more S*ub Design Goals* (*SDGs*) which are combined in definite ways to achieve the particular *DG* using e.g. a time based or a state based representation. Any *SDG* might be further broken down into more *SDGs* as needed. I.e. the abstraction level of the *SDG* is left to choice of the architects. A *Design Alternative* or *DA* is a possible architectural solution that fulfils the *DG*. Each *CFA* is analysed separately and if the existing *element*s cannot fulfil the chosen *DG*, one or more *DAs* that enable it to do so are assigned to the *CFA*. A *processed CFA* can be linked to zero *DAs* only if *elements* under consideration fulfil the *DGs*.

The abstraction of "Define process parameters" in Fig. 1 refers to deciding the *elements*, failure modes, DGs and generating the *CFAs*.

*Selection* as a noun is the subset of *DAs*, out of the set of all possible *DAs*, chosen to be included as part of the refined architecture. A *selection* is made as part of the concluding activity of the ATRIUM iteration i.e. in the "generate revised architecture" abstraction in Fig 1.

A *link* is the term used to describe a connection made between informational entities for the purposes of traceability of information. *Links* found in the *perceived certain* domain correspond to connections between *assumptions* to *CFAs*, *CFAs* to *DAs* and *DAs* to *selections*. Thus, *selections* can thus always be back-traced to *assumptions* at any point in the process.

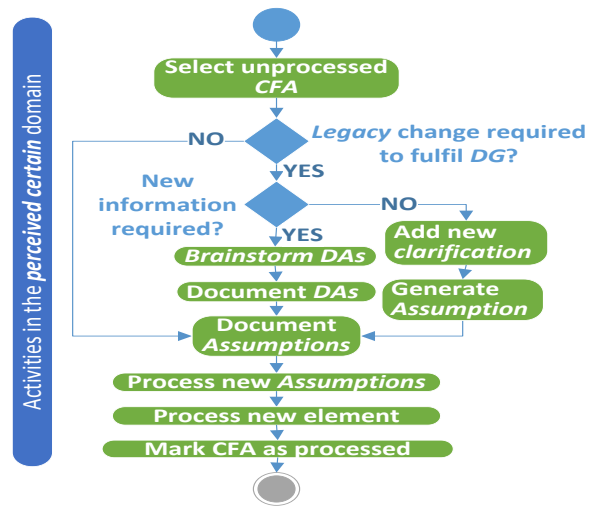

Fig. 2. Activities in the perceived certain domain

Fig 2 shows the activities of the process in the *perceived certain* domain and the management of *CFAs*. An *analysis* is performed per *CFA* and results in zero or more *DAs* while *assumptions* and *clarifications* are documented along with the appropriate *links*. When a *selection* is made, the reasons for rejecting the other *DAs* are recorded along with the reason for choosing the *selection*.

Fig 3 shows the process flow in the *uncertain* domain. *Links* are also found in the *uncertain* domain and are established from a *clarification* to an *assumption,* or a *task* to an *assumption*. A *clarification* contains detailed questions that must be answered by the secondary users like the experts. All necessary, but uncertain information become *clarifications* and are tracked individually. Each *clarification* requires the creation of an *assumption,* made based on information available and the judgement of the architects to allow work from the architects to progress in the *perceived certain* domain. If even the expert does not have access to the information immediately, the *clarification* becomes a *task.* The abstraction "convert clarification to task" in section 4.3 refers to this process. A *task* has resources allocated to it and is under the expert's responsibility to complete by a mutually agreed date. The expected dates for completion, name of architect responsible and expert are all documented with the *task*. Unfinished tasks at the end of an iteration of the process become *risks* and are collected in the *risk* list deliverable. This is depicted by the abstraction of "generate risk list" in Fig 1.

A *clarification* and an *assumption* are never deleted. *Clarification* or *tasks* (when complete) are marked as *resolved* only after *expert* consultation (i.e. if the *assumption* linked to the *clarification* was correct or if the existing *assumption* is marked as invalid and a new *assumption* is added and linked to both the relevant *CFAs* and the resolved *clarification)*. The link to the piece of information used for a particular *selection* is thus never lost. The ISO 26262 change management process which requires a change request, an analysis, documentation of decision, rationale and implemented changes is thus fulfilled at this abstraction level.

The process iteration is completed when (i) there are no *clarifications* remaining. (ii) there are no *unprocessed CFAs* and (iii) a *selections* has been made. A *selection* is chosen as part of the "Generate Revised Architecture" abstraction. The reason for allowing the *selection* to be made only at the end of the iteration is because *DAs* may potentially satisfy multiple *CFAs*, thus limiting the number of changes needed for the system as a whole. *ATRIUM* does not prescribe any particular method for the *selection* and allows organizations to use their own methods such as ATAM [9].

The abstraction of "process new *element*" refers to adding new *elements* and populating of the *CFAs* of those *elements*.

The abstraction of "process new assumption" refers to not only adding the *assumption* to the list but also the action of reviewing of all *CFAs* processed so far to see if the *assumption* has necessitated a new analysis of any *CFAs*. When new information is obtained that changes or invalidates an existing *assumption*, all linked *CFAs* are marked as *unprocessed* for subsequent analysis.

The deliverables of *ATRIUM* is the PAA comprised of the refined Preliminary Architecture (PA), the assumption list and a *risk* list. The way to interpret the results of *ATRIUM* is that the refined PA is valid under the assumptions listed in the *assumption* list and is subject to *risks* documented in the *risk* list.

*ATRIUM*, in addition to providing a PAA also documents *links* between failures and *assumptions*, thus providing easy access to rationales from previous iterations. This enables a basis for consistent discussion, and relevant decisions can be made accordingly and with documented justification.

## V. CASE STUDY

This section describes the application of *ATRIUM* in a case study at Scania's conducted by the authors of the paper. The important findings are reported, followed by a short discussion. The validity of the results and findings are discussed in Section 6.

### A. System Description

For the case study, *ATRIUM* was applied to a project in its early stages involving L3 level of automation [5] (conditional automation with a defined handover time before a human is responsible for the driving) in terms of functionality in a highway setting. The function was defined using the classification system introduced by the adaptive project in [23]. The specific parameter set used for this case study is available in full in appendix I. The existing architecture of a vehicle, augmented with changes from the technology roadmap of Scania was chosen as the input PA. From a practical point of view, the abstraction level of the platform *elements* was such that the evaluation of *ATRIUM* would be feasible in a reasonable amount of time, while still not being trivial. The communication paths consisted of multiple CAN busses, with a maximum bandwidth of 500 kb/s, multiple automotive grade Ethernet lines and of sensor/actuator specific cabling. The architecture comprised of all *elements* required in a modern state-of-the-art truck in the application scenario including multiple sensors such as cameras and radars, processing units Electronic Control Units (ECUs) etc. Actuators such as brakes and gearboxes were abstracted under their functional *elements*. External communication to the infrastructure and other vehicles was included under the abstraction of telematics.

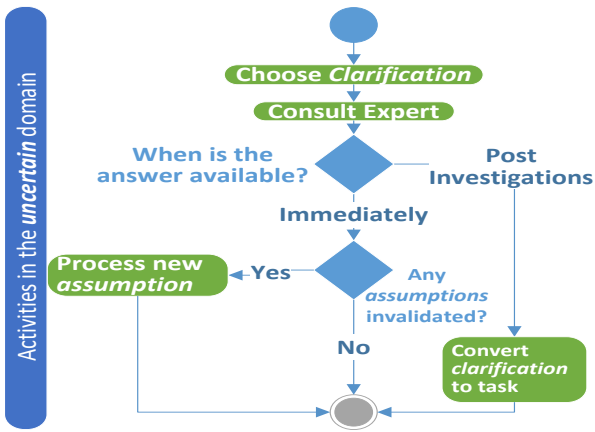

Fig. 3. Activities in the uncertain domain

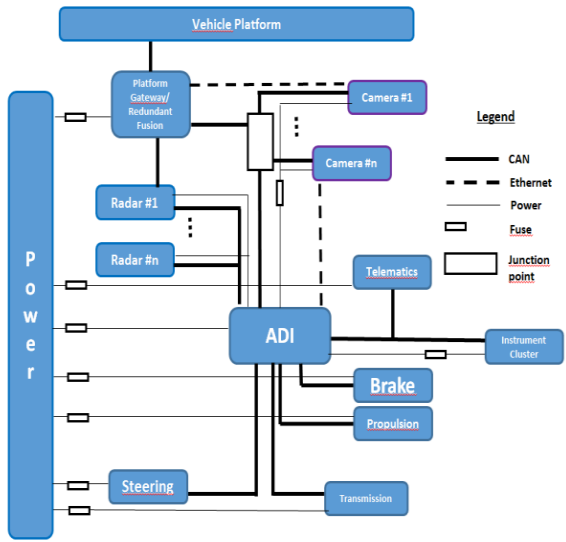

Fig. 4. Representative case study system architecture

For the sake of confidentiality, only a partial representative schematic is shown in Fig. 4 to provide an idea of the architecture that was used. The ADI node in the schematic specifically refers to Autonomous Driving Intelligence as defined in [10] and roughly corresponds to the intelligence needed to replace the human for automation in driving.

### B. Applying ATRIUM

In Step-group1, the first steps involved gathering inputs and then generating process parameters. The inputs were the system architecture described in section 5.1 as the unrefined PA with the high-level requirements of the function. The step of "deciding process parameters" involved deciding the *DGs* and scoping of failure modes. As a practical aspect, we decided to consider only single point failures, i.e. no two independent failures will appear simultaneously, to evaluate *ATRIUM* in a feasible amount of time.

We defined a single DG for the vehicle that needed to be achieved in case of a fault. The breakdown of *DGs* to *SDGs* was inspired by SAE J3016's definition of automation level 3, which allows the system to transition to a "minimal risk condition" at the end of a "handover time" if the driver has not responded to any request for takeover. An appropriate "handover time" was defined and documented and *DG* 1 was broken down into S*DG* 1 applicable for the duration the handover time and S*DG* 2 which would be applicable if the handover time had elapsed without driver intervention. *SDG* 1 would lead to a gradual decrease in speed and increase in distance from surrounding vehicles while *SDG* 2 would bring the vehicle to stop in the same lane. As an effect of these decisions, the vehicle required to have some level of basic vision, braking and steering available at all times i.e. even with a single point failure. Essentially, this part of *ATRIUM* helped deciding the ambition of the function's capabilities and extent of responsibility that Scania decided to take and realistic goals for the system to fulfil. *Elements* not needed to fulfil the functionality or the *DG* were removed from consideration and the number of *elements* was reduced from the entire platform to about 25 after this step. These were hierarchically segmented into four main segments with sub-segments encapsulating *Elements* with only local communication requirements. New *Elements* were added and some removed as the analysis progressed and more information was obtained. The considered failure modes in this case study were (1) full functional failure i.e. omission (2) loss of power and (3) communication failure per segment. *CFAs* were generated based on these parameters and DG1 was linked to them.

In step-group2, according to the process activity flow, each *CFA* was analysed and the functional effect on the vehicle under the fault was determined and recorded in a template similar to that of FMEA. If the baseline architecture was unable to achieve *DG1* given a *CFA*, a number of *DA*s were determined by the architects and linked to that *CFA*.

For example, given the *CFA* corresponding to a gateway *element* and a "communications failure", one functional effect is that all messaging between the two segments stops possibly causing the *SDG 2* ( reaching a safe stop) to become unachievable. Two possible *DAs* might then be to add a redundant communication path allowing *Elements* necessary for safe stop to communicate, or implementation of rudimentary safe stop functionality in each segment. Evaluation of the proposals was deferred to step-group4. Continuing this example, the relocation of *Elements* to a new segment while theoretically possible, depends in reality on the availability of network usage on that segment and a commitment from the *expert* on network design. This commitment was tracked as a *clarification* and later as a *task* when it emerged that the expert needed to perform several tests under specific network load conditions to investigate the result.

Thus a bank of assumptions, limitations, failure effects and questions along with architectural change proposals to the legacy architecture were assembled that would allow implementation of the desired behaviour after a failure. We came up with about 80 *CFAs* of which 9 produced *DAs*. Out of all these *CFAs*, 5 *CFAs* had more than one *DA*.

In step-group3, performed in parallel with Step-group2, *clarifications* were systematically resolved or converted to *tasks*. By the end of the first iteration, we had collected about 40 clarifications of which 25 had been resolved and about 10 converted to tasks. Approximately 30% of the clarifications and tasks invalidated assumptions and required rework. The tasks that were not completed became part of the *risk* list output and helped define the expected stability of the architecture.

In step-group4, the concluding activity of the iteration i.e. the harmonization of the *DAs* in terms of finding a *selection* of these with respect to a given set of metrics (specific to organization) was performed. The *selection* was judged to solve the single point failure shortcomings of the initial *PA* under the *assumptions* in the *assumption* list, while at the same time maintaining links between assumptions, failures and proposed solutions.

*C. Discussion*

As the case study task progressed, new limitations to preserve the applicability of the defined *DGs,* questions and implicit assumptions were discovered that affected behavioural and architectural choices. The flexibility of making decisions with incomplete information via *assumptions* and being able to track them via *clarifications* proved to be a strength with *ATRIUM* allowing architects to work relatively independently and in parallel with experts and safety engineers. We encountered a need for information that would require time-consuming testing activities several times during the process. Each time we were able to continue our work by using the best effort estimate from the expert while allowing them to continue detailed investigations.

It became clear from the beginning that having a legacy architecture helped constrain the design space by constraining the degrees of freedom and number of decisions that had to be made in the process. Though not necessary, having a legacy baseline architecture, made the execution of *ATRIUM* faster. If *ATRIUM* had been applied without legacy constraints, the resulting PA would probably be different due to the larger design space available. However, the refined PA obtained without consideration of legacy would be unrealistic for an established OEM, this PA could be still be of use to a new entrant to the market with no legacy considerations.

Many implicit assumptions that earlier would not have been documented were made explicit. We also note that this was further expedited by the presence of a relatively new colleague to the architecting group at Scania. It is thus recommended to choose a group makeup of people with different levels of experience and varied expertise to ensure less knowledge is tacitly assumed. *ATRIUM* also induced a level of detail in the way of working expected from safety critical systems and allowed for the documentation of the dependencies between different items and functions of the vehicle in crisp fashion.

The intensive nature of this process meant that there was a significant amount of work for the first iteration but each iteration of *ATRIUM* required less work from the architects and simultaneously allowed safety engineers to focus their efforts better and prioritize. The consistent discussions enabled by the application of *ATRIUM* was also instrumental in defining points of weaknesses in the functional requirements such as the lack of clarity of the responsibility split between driver and in the lack of details in the FSRs. These weaknesses made explicit by *ATRIUM* served as a basis for discussions between architects, safety engineers and experts and lead to a common understanding.

No new training was needed for the application of *ATRIUM,* as the techniques used such as FMEA are already well known in the industry. By the documentation of rejected choices and the rationale for rejecting them, *ATRIUM* allowed for selection of previously discarded choices upon change in the known information.

Overall, our opinion is that this process enabled risk reduction very early in the lifecycle of the development of the function and was very suitable for design using legacy *elements*. Expert judgement is still needed as part of *ATRIUM* but the tracking of links between information to architectural *elements* reduced the reliance on pure expert judgements and opinions. Along with the fact that failure modes and effects anchor discussions and reduce ambiguity, this linkage added repeatability to architectural decisions and smaller, faster impact analysis. Decisions that would have been made further down in the development cycle were made in advance and were documented better leading to faster development times overall.

*ATRIUM* embraced the experience of the architects while not compromising on rigour and provided useful results as judged by all of its primary and secondary users. To the best of the authors' knowledge and from an initial literature review, *ATRIUM* is the only process that addresses the engineering of the PAA. *ATRIUM* was judged after the case study to have satisfied the requirements (Section 4.1) used to build it and made headway into reducing the challenges described in the analysis of the problem (Section 3). *ATRIUM* has thus been institutionalized in at Scania as a standard way of working with safety critical systems involving high automation.

VI. EVALUATION AND LIMITATIONS

This section gives a short explanation of the evaluation methods used.

The lack of metrics of the process that was replaced i.e. consensus by experts, and the exploratory nature of the work, necessitated qualitative research in the design of *ATRIUM*. As such, it suffers from the weaknesses of qualitative research in validation. This has been reduced to some extent by the use of triangulation [11] with (i) a review of the literature, (ii) multiple workshops with varying participants used for evaluation and (iii) evaluating the quality of the results of the case study by comparison to that of a control group using traditional methods.

The process itself was evaluated in multiple workshops with the primary users (architects not involved in the design of *ATRIUM*) and revised until it was judged easy to use, applicable to the class of the problem i.e. fit for purpose, and to implement the requirements section 5.1. The secondary users of *ATRIUM* i.e. safety engineers and experts were part of different workshops and judged *ATRIUM* to facilitate their work with functional safety while providing a suitable separation of their concerns from those of the architects. The safety engineers found the PAA to be of better quality than what had been generated by traditional methods. Thus, ATRIUM was judged to be fit for purpose and better than prevailing methods by both the primary and secondary users.

The results obtained from the case study were evaluated separately by comparison with results obtained from a control group of architects who analyzed the same function. The control group's results were found to be comparable to that of *ATRIUM*, except for the lack in the documentation and traceability.

The results obtained in the case study in Section 5 further validate *ATRIUM*' applicability and it is currently seen to be essential for functions with high automation within Scania. due to (i) the diversity in the number of *elements* and (ii) the degree of uncertainty due to the complexity of the function itself. This level of rigour may not be needed for functions of a simpler nature. Even so, since *ATRIUM* has only been evaluated in a single function, its applicability needs to be validated by further case studies. This is targeted as future work.

Care was taken in the design and review of *ATRIUM* to ensure that it was generalizable in that there were no organization specific dependencies in any of the activities.

## VII. RELATED WORK

This section focusses on the comparison of the PAA or equivalent in safety literature and the effect of the lack of it in the automotive domain.

Interestingly, though the ISO 26262 does not place any requirements or restrictions on the PAA, the functional safety standard for aviation systems development i.e. ARP4754A [12] does explicitly define the need and requires tracking of assumptions made during architectural decisions. Leveson in her work with intent specifications [13] also discusses the significance of documenting assumptions for operational safety. Contracts based design, which has been proposed for design of safety critical system in automotive in papers such as [14] are directly benefitted by reliably keeping track of assumptions.

The effect of the lack of guidelines and the ambiguous definitions used for the PAA is evident, with a short look into the academic and industrial papers from the automotive domain, in how the PAA has been addressed. While we found many papers that evaluated the use of the ISO 26262 standard on examples or on industrial case studies, no particular paper expanded upon how their PAA was obtained. The PAAs used in literature were found to be of varying levels of detail and content. Taylor et al. in [15] uses a hardware inspired PAA, Westman and Nyberg in [16] use a pure software based PAA, identify *elements* as software *elements* and leverage the information from their legacy example to make their case. [17] also includes the mechanical considerations such as installation space etc. [18] and [19] address the PAA as functional blocks making no direct reference to a particular implementation technology. Thus, depending on what type of legacy systems are used or what is more important to an organization etc., different definitions of the content of the PAA may exist. The definition of the PAA is also discussed in [20] where commonalities between ISO 26262, IEC 61508 and ASPICE are derived. Tagliabó et al equate the purpose of the PAA to the EAST-ADL analysis architecture [21]. There is clearly much confusion about the definition and the contents of the PAA was one reason why *ATRIUM* does not place restrictions on the type of *elements* used.

The authors agree with the views expressed in [22] in the importance of early safety evaluation and see incomplete PAAs as a major obstacle in this regard.

## VIII. CONCLUSIONS AND FUTURE WORK

*ATRIUM* solves an immediate need in the automotive industry where the structure demanded by ISO 26262, must be grounded in the reality that decisions need to be made under uncertain information.

Replacing the previous technique of designing the PAA i.e. consensus amongst experts, it introduces traceability from an uncertain piece of information to a definite architectural *element* in the PA by establishing a framework to handle uncertainty in architectural decisions. The enforced traceability ensures that there is a more focused and easier impact analysis on changed preconditions. The rationale management and assumption management that is imposed allows for guidance for such an impact analysis and as a reference for other architects, thus enabling repeatability and consistency in both decision-making and discussion.

The deliverables of *ATRIUM* including the preliminary architecture, the related assumptions bank and *risk* list help the safety engineers and project managers by providing them with realistic information about architectural *risks*.

*ATRIUM* was applied at one of Scania's functions involving high automation and was found to be useful enough to be institutionalized as a standard way of working within the organization. *ATRIUM* is planned to be applied to a function of L4 level of automation and the results will be used to further improve the process.

The work so far has been accomplished using standard office software and considerable manual effort. Abstraction of the data managed by *ATRIUM* into a metamodel to enable tool support is a prioritized next step.

It was noted that there are classes of information that are needed by architects that are common to different subsystems in the vehicle. The efficiency of *ATRIUM* could be greatly improved by having information e.g. from diagnostic monitors available while making architectural decisions at a suitable abstraction.

Finally, the generalizability of *ATRIUM* and its benefits were appreciated throughout its successful application by its users. *ATRIUM* is hence planned to be used in other domains than just automotive to further evolve it towards domain independence.

APPENDIX I

CLASSIFICATION OF FUNCTION ACCORDING TO THE ADAPTIVE SYSTEM

| AdaptIVe parameter | Name | Value |
|---|---|---|
| 1.1 | Type | Truck |
| 1.2.1 | Time to collision | Large |
| 1.2.2 | Duration | Continuous |
| 1.2.3 | Automation | Level 3 conditional automation |
| 1.2.4 | Speed Range | Low, Medium and high |
| 1.2.5 | Control force | Low, Mid |
| 1.2.6 | Time headway | Standard |
| 1.2.7 | Trigger | 1.2.7.1 System initiated |
| 1.2.8 | Coordination | 1.2.8.2 Without coordination |
| 2.1 | Driver qualification | 2.1.2 Professional |
| 2.2 | Driver Location | 2.2.1 Inside vehicle |
| 2.3 | Driver's Monitoring task | 2.3.2 Need not monitor |
| 2.4 | Driver activation | 2.4.1 Attentive 2.4.2 Inattentive |
| 2.5 | Driver is capable to control his vehicle | 2.5.2 Yes |
| 3.1.1 | Traffic mixed | 2.5.2 Yes |
| 3.1.2 | Traffic participants | 3.1.1.1 Yes |
| 3.1.3 | Traffic flow | 3.1.2.3 Motorized, type B |
| 3.2.1 | Road type | N/A |
| 3.2.2 | Road accessibility | 3.2.2.1 Public 3.2.2.2 Private |
| 3.2.3 | Road condition | 3.2.3.1 Good 3.2.3.2 Slippery 3.2.3.3 Bumpy |
| 3.2.4 | Road geometry | 3.2.4.1 Straight 3.2.4.2 Curved 3.2.4.3 Steep |
| 3.2.5 | Road infrastructure | 3.2.5.1 Physical cut-off 3.2.5.2 Good lane markings 3.2.5.3 Guard rails 3.2.5.5 Emergency lanes |
| 3.3.1 | Good visibility | 3.3.1.1 Good Visibility |
| 3.3.2 | Poor visibility due to obstacles | 3.3.2.1Vehicles |
| 3.3.3 | Poor visibility due to weather conditions | N/A |

ACKNOWLEDGMENT

Support from FFI, the Swedish National funding agency for Vehicle Strategic Research and Innovation and Vinnova through the ARCHER (proj. No. 2014-06260) is acknowledged. Ola Bergqvist from Scania was instrumental in this work, both in setting up this task and through his guidance. We would also like to thank Anna Beckman and Magnus Gille from Scania for their efforts in reviewing the text of this paper.